\documentclass[twocolumn, english, aps,reprint, superscriptaddress,showpacs,longbibliography,showkeys,prl]{revtex4-2}
\usepackage{amsmath,amssymb,bbm,mathrsfs,bm,braket,color,graphicx,comment,xcolor,dsfont,multirow}
\usepackage[colorlinks,citecolor=blue,urlcolor=blue,linkcolor = blue]{hyperref}

\begin{document}
\date{\today}

\title{Fault-tolerant multi-qubit gates in Parity Codes}
\author{Anette Messinger}
\affiliation{Parity Quantum Computing GmbH, A-6020 Innsbruck, Austria}
\author{Christophe Goeller}
\affiliation{Parity Quantum Computing France SAS, 75016 Paris, France}
\affiliation{Parity Quantum Computing Germany GmbH, D-20095 Hamburg, Germany} 
\author{Wolfgang Lechner}
\affiliation{Parity Quantum Computing GmbH, A-6020 Innsbruck, Austria}
\affiliation{Parity Quantum Computing France SAS, 75016 Paris, France}
\affiliation{Parity Quantum Computing Germany GmbH, D-20095 Hamburg, Germany} 
\affiliation{Institute for Theoretical Physics, University of Innsbruck, A-6020 Innsbruck, Austria}

\begin{abstract}
We present a set of efficiently implementable logical multi-qubit gates in concatenated quantum error correction codes using parity qubits. In particular, we show how fault-tolerant high-weight rotation gates of arbitrary angle can be implemented on single physical qubits of a classical stabilizer code, or on localized regions of full quantum error correction codes. Similarly, we show how transversal CNOT gates can implement logical parity-controlled-NOT operations between arbitrarily many logical qubits. Both operation types can be implemented and in many cases parallelized without the use of lattice surgery or the need for complicated routing operations.
\end{abstract}

\maketitle

\paragraph{Introduction}
Quantum computers promise great advances in areas such as chemistry, materials science, and cryptography, offering speedups beyond what is classically possible \cite{dalzell2023quantum}. However, quantum information is intrinsically fragile and the noise affecting qubits severely limits the depth and scale of quantum algorithms. 

Over the past decades, enormous effort has gone into developing robust quantum memories using quantum error correction (QEC) codes, encoding the logical information redundantly in many physical qubits \cite{Nielsen2011}. The dominant approach so far has relied on planar codes such as the surface code \cite{fowler2009high,google2025quantum}, requiring only nearest-neighbour interactions at the cost of quadratic qubits overhead. Although recent efforts on more efficient quantum low-density parity-check (QLDPC) codes, such as the bivariate bicycle codes \cite{bravyi2024high}, promise substantial reductions in qubit count while preserving some local structure, efficiently realising universal fault-tolerant logical gates within these codes is an ongoing research problem \cite{krishna2021fault,cohen2022low,breuckmann2024fold,malcolm2025computing,yoder2025tour}. Additionally, even for approaches making use of lattice surgery \cite{horsman2012surface,fowler2018low,litinski2019game}, such as the surface code, parallelizing logical operations while keeping the qubit overhead under control remains a difficult yet important challenge \cite{herr2017optimization}.

Parity codes \cite{Fellner2022universal,messinger2023} aim to address this problem by enabling logical long-range entangling gates using solely local operations.
As instances of classical linear codes, they provide intrinsic protection against one type of error, and can be concatenated, for example, with asymmetric QEC codes or implemented on noise-biased qubits \cite{messinger2024}. The qubits added in such codes act to increase redundancy and enable logical connectivity at the same time.
Importantly, keeping the redundancy of the parity code in a single basis allows one to address the same logical qubit on different physical qubits at the same time without violating the no-cloning theorem.
The resulting parallelizable logical multi-qubit gates offer a promising route toward scalable quantum computing with practical resource requirements.

The remainder of this work is organized as follow. We start by introducing the parity code and its labelling formalism followed by a detailed explanation of the fault-tolerant implementation of the parity-controlled NOT gate, as well as many-body rotations of arbitrary angle.

\paragraph{The Parity Code}

We use the term parity code to describe a classical code whose stabilizers are chosen to obtain a certain physical-to-logical mapping $P_k \rightarrow \bar P_i \bar P_j ...$ where $\bar P$ represents a logical operator. This allows one to easily access logical multi-qubit operators by directly acting on single physical qubits. In this work, we consider $P=Z$ but the same holds for any other basis. In principle, any classical stabilizer code can be described as a parity code. 

\begin{figure}
    \centering
    \includegraphics[width=.8\linewidth]{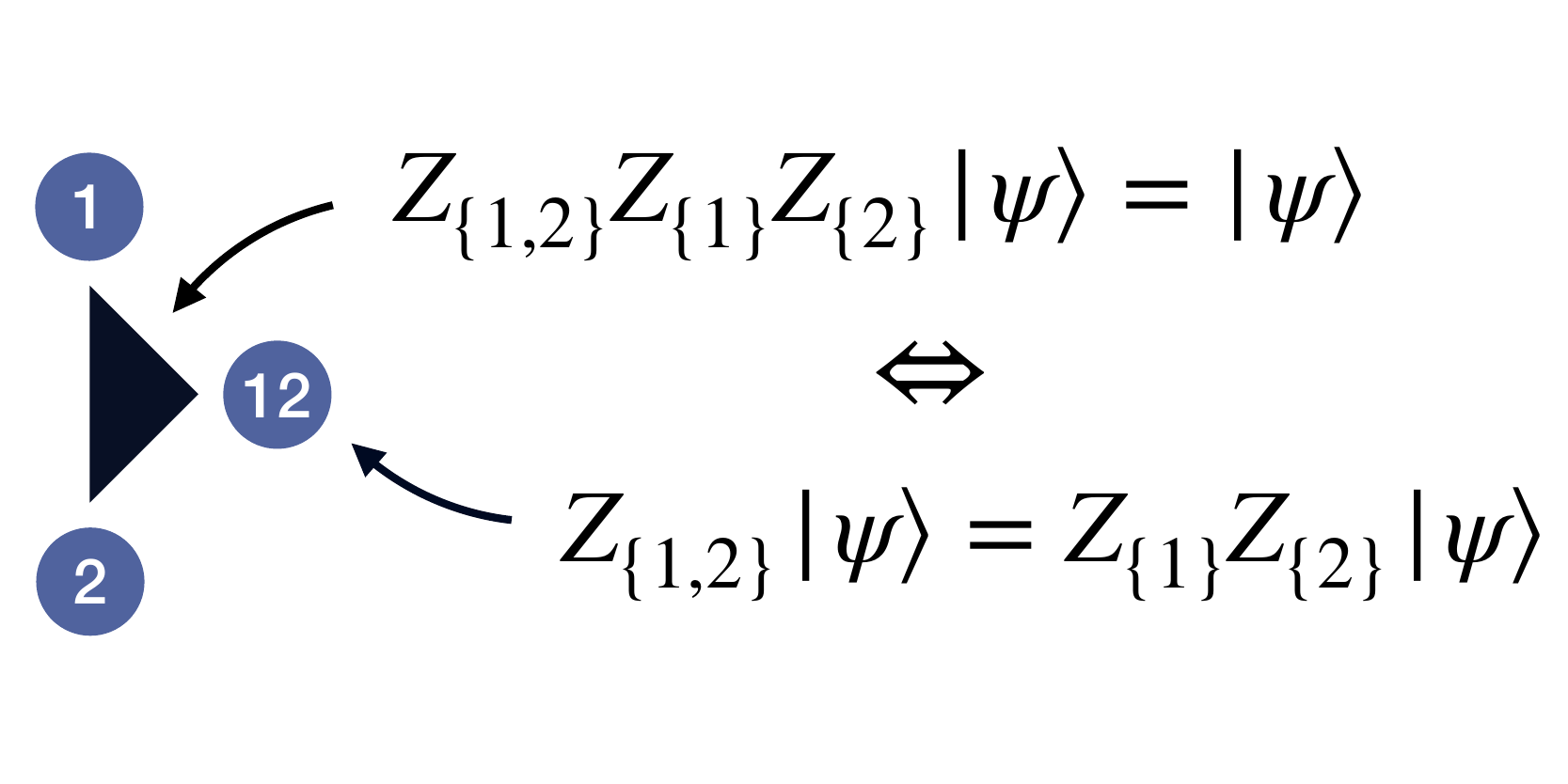}
    \caption{Equivalence of stabilizer condition and relation between physical $Z$ operators. If two physical qubits have already been assigned a logical operator (indicated by their label), in this case $Z_{\{1\}}=\bar Z_1$ and $Z_{\{2\}}=\bar Z_2$, then the third qubit must have a mapping to their product  and we label it accordingly, $Z_{\{1,2\}} =Z_{\{1\}}Z_{\{2\}}= \bar Z_1 \bar Z_2$.}
    \label{fig:stab-label}
\end{figure}

One method to obtain the available mappings in a given classical code is to reformulate the stabilizer conditions to relate certain physical $Z$ operators to products of others and ultimately to products of logical $Z$ operators. One can always start by defining direct mappings from physical to logical operators as $Z_{\{i\}} = \bar Z_i$ for some qubits \footnote{Note that we use the equality symbol to denote equality up to stabilizers.}. Whenever all but one qubit of a stabilizer have been assigned such a mapping, the mapping of the remaining qubit follows from the stabilizer. For example, if we have assigned two qubits of a three-body stabilizer  $Z_{\{1\}}=\bar Z_1$ and $Z_{\{2\}}=\bar Z_2$, then the third qubit must have a mapping to $\bar Z_1 \bar Z_2$ and thus we give it the label $Z_{\{1,2\}}$, to satisfy the stabilizer condition 
\[Z_{\{1,2\}}Z_{\{1\}}Z_{\{2\}}|\psi\rangle = \bar Z_1 \bar Z_2 \bar Z_1 \bar Z_2 |\psi\rangle = + |\psi\rangle,\]
as depicted Figure \ref{fig:stab-label}.

Here and in the following we use the convention to label physical qubits of a code with the set of indices of the logical qubit operators they have a mapping to (cf. parity flow formalism \cite{Klaver2025}, shown for the same example in the end matter, for a more extended form of this labelling convention). We call physical qubits whose label contains more than one logical qubit index \textit{parity qubits} and single-labelled qubits \textit{base qubits}.
Note that the number of independent labels is given by the number of logical qubits in the code.

From these labels, one can also easily read off the definition of logical $\bar X$ operators as the product of all physical $X$  operators on qubits whose label contain the corresponding logical qubit index, $\bar X_i = \prod_{\{\mathcal{L} |i \in \mathcal{L}\}} X_\mathcal{L}$ (this anti-commutes with exactly each physical $Z$ whose logical effect contains $\bar Z_i$).

Instead of just labelling the physical qubits of existing stabilizer codes to find the available parity qubits, it is often useful to construct the code directly with the desired logical connectivity in mind, i.e., with the set of parity qubits it should contain. The most common example for such a tailored code is the [$k(k+1)/2$, $k$, $k$] LHZ layout \cite{Lechner2015, Fellner2022universal}, where $k$ is the number of logical qubits. This LDPC code family was originally introduced to solve optimization problems of all-to-all connected problem graphs using one parity qubit for each possible two-body term.

It was later shown that even for mappings to products of more than two logical operators , {$Z_{\{i,j,k,...\}} = \bar Z_i \bar Z_j \bar Z_k ...$}, one can always find a code layout supporting the desired parity qubits with local stabilizer operators of order four or less on a square lattice \cite{terhoeven2023}.

Adding or removing parity qubits to or from existing code layouts can be done via CNOT-based encoding and decoding circuits \cite{Fellner2022universal} or, in a more parallelizable manner, via measurement-based code deformation \cite{messinger2023}: A parity qubit with label $\mathcal{L}$ can be added to the code by adding a fresh qubit in the state $\ket +$ and measuring a stabilizer operator connecting this qubit to existing qubits whose labels combine to the desired new label. The measurement then enforces the desired relation between the qubits (if the stabilizer is measured as $-1$, either flip the new qubit or update the Pauli frame). A parity qubit can be removed from the code by simply removing one stabilizer which links it to the rest of the code, and measuring the qubit in $X$ (performing conditional $Z$-corrections or Pauli-frame updates on the remaining qubits of the removed stabilizer).  If the removed qubit was in multiple stabilizers, a new basis of the stabilizer operators must be chosen for the remaining code (e.g., multiplying all those stabilizers with the one removed stabilizer).\\

\paragraph{Fault-tolerant basic gate set}
In Ref.~\cite{messinger2024} a universal fault-tolerant gate set in concatenated parity codes was demonstrated. For a fully fault-tolerant implementation, the underlying qubits of the parity code must be protected against at least one type of error (in our example $X$ flips) and a bias-preserving CNOT gate needs to be available. These requirements can be fulfilled by simple repetition codes or asymmetric surface codes, but also some inherently noise-biased quantum computing platforms, like for example cat qubits \cite{webster2015, guillaud2021, aliferis2008, guillaud2019, chamberland2022}.
The demonstrated gate set contains transversal CNOT gates, but also $\bar R_{ZZ(...)}(\pi/2)$ and $\bar R_{ZZ(...)}(\pi/4)$ gates implemented via magic gate teleportation of S and T gates to parity qubits. In an implementation with cat qubits, dedicated protocols have been proposed \cite{mirrahimi2014dynamically,puri_bias-preserving_2020,chamberland2022,ruiz2025unfolded} which can be used for implementing also many other gates, like for example $\sqrt{X}$ , $H$, $Q=SHS$ and Toffoli gates fault-tolerantly or to perform more efficient magic distillation.\\

\paragraph{Parity-controlled-NOT}
In the parity code, physical CNOT gates never violate the code space on their control side. Therefore, any sequence of CNOT gates from the same control label $\mathcal{L}_c$, targeting all qubits of a logical operator $\bar X_i$,
\begin{equation}
    \prod_{\{\mathcal{L}_t | i\in \mathcal{L}_t\}} \text{CNOT}_{\mathcal{L}_c\mathcal{L}_t}
\end{equation}
is in principle an allowed logical operator, even if the control was a parity qubit. 
In that case, the operation can be interpreted as a logical parity-controlled NOT, i.e. the target flip is conditioned on the parity of the logical qubits indexed in the control label being odd. Such a gate is equivalent to a sequence of logical CNOT gates controlled on each of the logical qubits of the control label,
\begin{equation}
    \prod_{j\in \mathcal{L}_c}\overline{\rm{CNOT}}_{ji}.
\end{equation}
Note that, for a fault-tolerant use of this operation, the control parity qubit needs to be sufficiently protected and match the protection of the logical target qubit: If the logical target's $\bar X$ operator is spread over $d$ code qubits, a transversal implementation of the CNOT gate requires the use of $d$ separate control parity qubits (each with the same label and thus containing the same $Z$-information), such that each physical CNOT can be controlled from a separate qubit \footnote{Depending on the underlying qubits, an alternative could be to do the CNOT gate non-transversal and interleave it with syndrome extraction rounds to prevent error spreading.}. 

In a combination of the parity code with an (asymmetric) surface code, the target qubit may also not be protected by the parity code at all, but solely from the underlying surface code, in a more symmetric form. This can for example be the case when using traditional magic state factories to prepare magic resource states. In such a case, a single control parity qubit is sufficient, but its underlying encoding must be at least temporarily matched in distance (both $X$  and $Z$) to the target's encoding.\\

\paragraph{Many-body rotations}
It has been shown already in \cite{messinger2024} that gate teleportation of S or T gates directly onto parity qubits enable the implementation of logical $\bar{R}_{ZZ...}$  gates of angles $\pi/2$  or $\pi/4$,  respectively. For most platforms, efficient distillation and teleportation schemes only exist for these angles (or a slightly larger set), and thus arbitrary logical gates $\bar{R}_{ZZ...}(\alpha)$ cannot be constructed in general from only those $Z$-type rotations. 

\begin{figure*}
    \centering
    \includegraphics[width=1\textwidth]{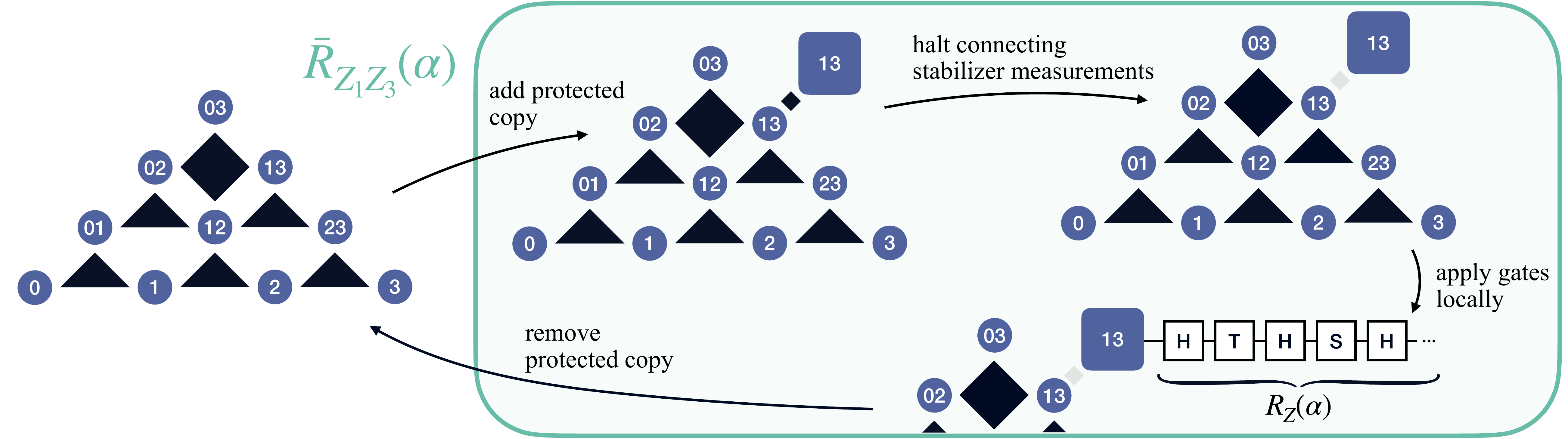}
    \caption{Schematic implementation of a fault-tolerant multi-qubit rotation on the example of $\bar R_{Z_1 Z_3}(\alpha)$ in the LHZ layout. First, a second instance of parity qubit $13$ is added to the code (blue square). This ``copy'' should be protected on its own (e.g. by encoding it within another classical code, or by increasing the Z-distance of the existing underlying encoding) and not rely on the remaining parity code anymore. Multiple rounds of stabilizer measurements might be needed to ensure the protection. Once the parity qubit copy is sufficiently protected, the stabilizer connecting it to the rest of the parity code can be excluded from further syndrome extraction rounds and a decomposition of the rotation $R_Z (\alpha)$ be applied to the copy (fault-tolerantly regarding it's own encoding). After the decomposition is complete, the connecting stabilizer to the parity code can be activated again or the copy be removed directly.} 
    \label{fig:ParityR-prot}
\end{figure*}

In the following we show that we can still use parity qubits to construct such arbitrary angle multi-qubit rotations by using also H gates directly on parity qubits (see Fig. \ref{fig:ParityR-prot}), allowing us to still use the standard Clifford + T decomposition for universal computation \cite{bravyi2005}.

Even though the gates used for those arbitrary angle decompositions are not fault-tolerant or even code-preserving in the parity code, we note that any stabilizer violations are only temporary. The final effect of the composed rotation gate will be in the Z basis again and thus one should end up back in the code space, at least approximately, after a sufficiently good decomposition.
As the parity qubit of interest must be additionally protected for the magic state injection steps anyway, the same protection should also be kept up during implementation of the H gates or any other operations applied in the decomposition. For example, if this protection is done via an additional repetition encoding of the parity qubit, then the H must be implemented as a logical gate within this repetition code (while the rest of the parity code can be ignored at this point).

In order to not reduce the code distance of the remaining parity code, it can be beneficial to keep an additional ``copy''of the parity qubit of interest, i.e. use two parity qubits with the same labels \footnote{Note that repeating a qubit with the same label only copies the information in one basis, i.e. does not clone quantum information}. One of them can be additionally protected and host the decomposition of the rotation, while the other remains untouched by direct operations and can be used to keep up the syndrome extraction and error correction protocols in the rest of parity code. As long as all code-violating operations happen on the added copy, the only violated stabilizer will be the one connecting the two parity qubits. For the duration of the operation, we can thus simply ignore this stabilizer and exclude it from syndrome extraction - each side of it is well enough protected on its own.

The protocol for an arbitrary $\bar R_{ZZ(...)}(\alpha)$ rotation is then as follows (see Fig. \ref{fig:ParityR-prot} for an example of a two-body rotation):
\begin{enumerate}
    \item Add a fully protected copy of the parity qubit with the label of the desired rotation, perform syndrome extraction with the stabilizer connecting the parity qubit to its protected copy active until the protection of the copy can be trusted independently.
    \item Stop measuring the connecting stabilizer, keep all other syndrome extraction running.
    \item Perform a decomposition of $R_Z(\alpha)$ on the protected copy (as a logical operation within the code that protects the copy).
    \item Remove the protected copy (via measurement or other code deformation methods). This can be done even without an intermediate step of reactivating the connecting stabilizer operator as it is expected to be satisfied anyway after the decomposition and thus needs no additional measurements.
\end{enumerate}

With sufficient protection of the copied qubit and accuracy of the rotation decomposition, this protocol allows to perform arbitrary multi-qubit Pauli rotations fully fault-tolerantly. Note that since all operations are localized in a small region around a single parity qubit, many such operations can in principle be parallelized without routing issues - as long as enough space for the copies and access to magic resources are available.\\

\paragraph{Conclusion}
We have shown how to make use of parity qubits naturally occurring in every classical stabilizer code for more efficient logical computation. We stress in particular that codes can be tailored to host exactly the parity qubits needed to significantly speed up many quantum algorithms, especially in the fields of Hamiltonian simulations and optimization problems \cite{low2019hamiltonian,farhi2014quantum} where many operations in the same Pauli basis occur at the same time. The construction of explicit layouts and protocols for maximal parallelization and optimal utilization of magic state factories will be the content of future work.

We have demonstrated fully fault-tolerant implementations of all introduced operations using concatenations with other codes, e.g. surface codes or hardware-native encodings. 
Using code deformation, the parity codes can be used flexible and also combined with other schemes, like surface code lattice surgery, to fully optimize quantum algorithm implementations. We expect the introduced gates to be particularly interesting in cases where transversal gates are preferred to lattice surgery, like for example in recent proposals on operating neutral atom quantum computers fault-tolerantly \cite{Zhou2025, Zhou2025resource}.\\

\paragraph{Acknowledgements -}
This project was supported by FFG Funding (Project No. FO99918691) as part of the international Eureka cooperation as well as FFG Basisprogramm (Project No. FO999924030), and the Federal Ministry of Research, Technology and Space (BMFTR) program on Quantum Technologies – From Basic Research to Market (Contr. No. 13N17402). 
The authors further thank Nitica Sakharwade, Michael Schuler and Michael Fellner for helpful discussions.

\bibliographystyle{apsrev4-2}
%

\begin{figure}
    \centering
    \includegraphics[width=0.8\linewidth]{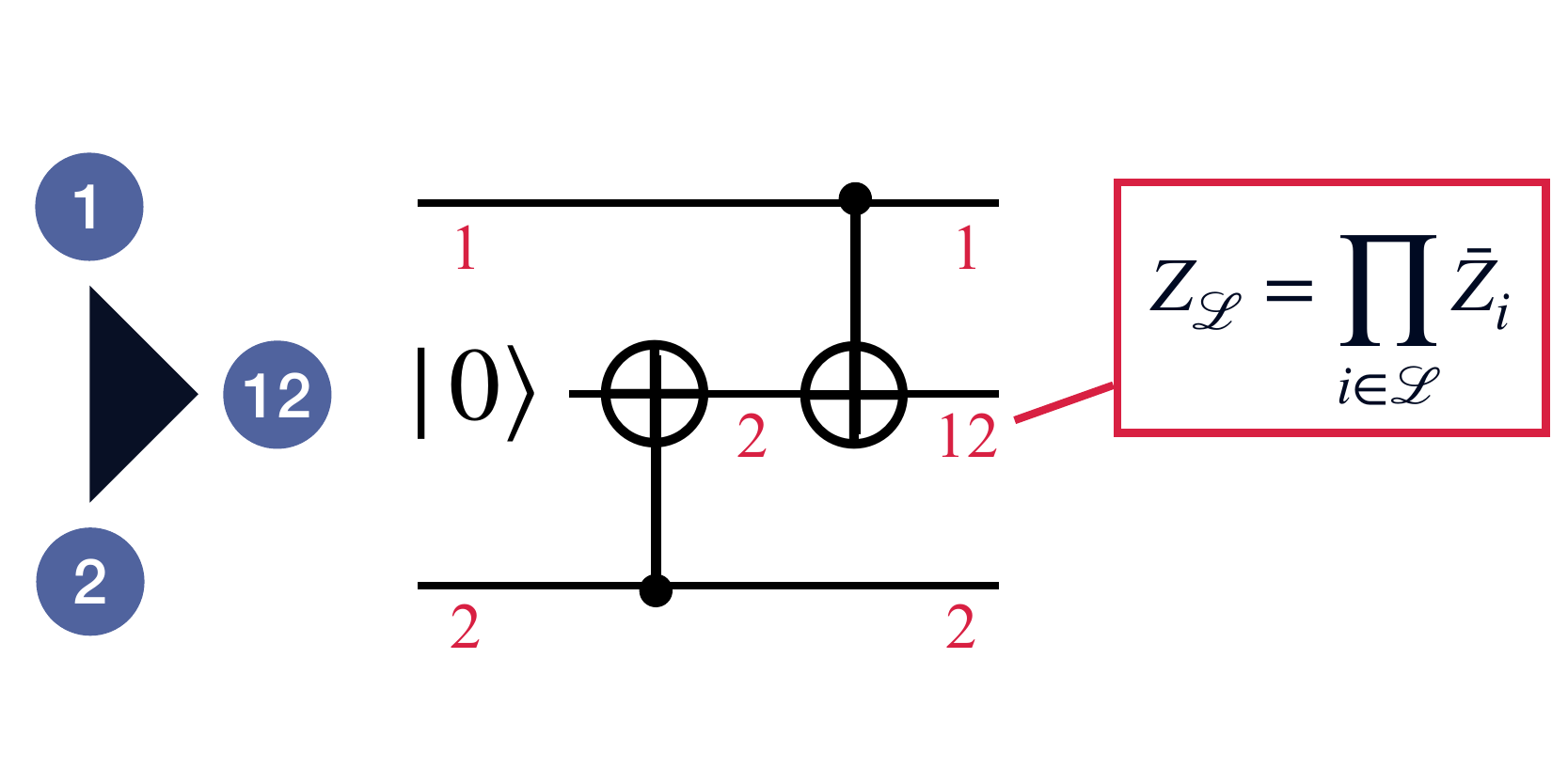}
    \caption{Example of physical-to-logical operator mapping for the encoding circuit of a three-body $Z$ stabilizer. After encoding, a physical $Z$ on the added qubit has the effect of a $Z$-product on the two logical qubits (labelled 1 and 2). The two original physical qubits retain a direct mapping to the corresponding logical qubits.}
    \label{fig:tracking-example}
\end{figure}

\appendix
\paragraph{End Matter - Parity Labels from Encoding Circuit}
If an encoding circuit for a given code is known, one can derive the valid qubit labels using the parity flow formalism as a circuit tracking method \cite{Klaver2025}. Before the encoding circuit is applied, we can associate each logical qubit with a physical qubit and the logical Pauli operators with the corresponding physical operators on those qubits. One can than track how the effect of physical qubit operators changes with the encoding circuit. A physical Pauli operation $P$  applied after an encoding circuit $C$ has the same effect as the operator $C^\dagger P C$ applied before the encoding circuit, i.e. applied on the logical qubits. For redundant encodings, one additionally needs to take into account the added qubits which were initialized in a certain state, and the fact that some Pauli operators have a trivial effect on this state. For classical stabilizer codes, we can assume that all those qubits are initialized in $|0\rangle$ and are thus invariant to Pauli-$Z$ operations. From this, one can derive relations which map each physical Pauli-$Z$ operator to either a single logical $\bar Z$ or a product of logical $\bar Z$ operators. We will illustrate this mapping in the following example, using the graphical short-hand notation of the flow formalism for a single basis to represent the mapping of physical to logical Pauli $Z$ operators.
The encoding circuit shown in Figure. \ref{fig:tracking-example}
encodes two logical qubits labelled 1 and 2 in three physical qubits with a three-body $Z$ stabilizer between them. In the encoded state, the third qubit will have a mapping $Z_{\{1,2\}} = \bar Z_1 \bar Z_2$. The two original physical qubits still have a direct mapping $Z_{\{1\}}=\bar Z_1$ and $Z_{\{2\}}=\bar Z_2$. 

\end{document}